\documentclass[conference]{IEEEtran}
\usepackage{cite}
\usepackage{amsmath,amssymb,amsfonts}
\usepackage{algorithmic}
\usepackage{graphicx}
\usepackage{textcomp}
\usepackage{xcolor}
\usepackage{url}
\usepackage[capitalize]{cleveref}
\usepackage{tlatex}
\def\BibTeX{{\rm B\kern-.05em{\sc i\kern-.025em b}\kern-.08em
    T\kern-.1667em\lower.7ex\hbox{E}\kern-.125emX}}
\usepackage{hyphenat}
\begin{document}

\title{Towards a Formal Verification of the Lightning Network with TLA\textsuperscript{+}\\
}

\author{\IEEEauthorblockN{Matthias Grundmann, Hannes Hartenstein}
    \IEEEauthorblockA{
    \textit{KASTEL Security Research Labs}\\
    \textit{Karlsruhe Institute of Technology (KIT)}\\ 
    Karlsruhe, Germany}
}

\maketitle

\begin{abstract}
Payment channel networks are an approach to improve the scalability of blockchain-based cryptocurrencies.
Because payment channel networks are used for transfer of financial value, their security in the presence of adversarial participants should be verified formally.
We formalize the protocol of the Lightning Network, a payment channel network built for Bitcoin, and show that the protocol fulfills the expected security properties.
As the state space of a specification consisting of multiple participants is too large for model checking, we formalize intermediate specifications and use a chain of refinements to validate the security properties where each refinement is justified either by model checking or by a pen-and-paper proof.
\end{abstract}


\section{Introduction}

Blockchain-based cryptocurrencies do not scale well with respect to their transaction throughput.
One approach to improve said scalability are Payment Channel Networks -- a second layer on top of a blockchain that processes transactions without writing each transaction to the blockchain.
A payment channel between two users is opened by performing one transaction on the underlying blockchain.
Once a payment channel is open, it allows for performing an unlimited number of transactions between its participating users without writing to the blockchain.
Finally, a payment channel is closed by publishing a second transaction on the blockchain.
In a payment channel network, the participating users are connected by payment channels and can perform multi-hop transactions so that the sender and the recipient of a transaction do not need to have a payment channel directly connecting them but it suffices that a path between sender and recipient over a set of payment channels exists.

The security model for payment channels requires that honest users cannot loose their funds even if all other users behave adversarially.
To avoid financial loss caused by design flaws in a payment channel protocol, it should be verified that the protocol is secure.
In this paper, we analyze the security of the protocol of the Lightning Network \cite{poon_bitcoin_2016}, a payment channel network for Bitcoin \cite{nakamoto_bitcoin:_2008}, for which different implementations exist and which is used in practice.
Our goal is to verify security properties of the Lightning Networks' protocol.
While this goal has already been approached in previous work \cite{kiayias_composable_2020}, we aim at verifying the properties in a largely machine-checked way because the complexity of the Lightning Network's protocol is difficult to handle.
Our general approach is to formalize the Lightning Network's protocol in TLA\textsuperscript{+} and verify using model checking that the protocol fulfills the security property that honest parties retrieve at least their correct balance.
Due to the complexity of the Lightning Networks' protocol, we cannot directly model check the protocol specification.
Instead, we use model checking for the most difficult proof steps and we provide pen-and-paper proofs to extend the statements about specifications that we can model check to the whole protocol specification.

To concretize, we formalize an ideal functionality of a payment channel that abstracts the behavior of the Lightning Network's protocol.
We show that the formalized protocol specification of a payment channel refines the ideal channel functionality by explicitly specifying a refinement mapping between the formalized protocol specification and the ideal channel functionality. We verify the validity of the refinement mapping using a model checker.
By using the ideal channel functionality, we specify an abstraction of the Lightning Network's protocol for multi-hop payments.
We use a model checker to verify that this specification of idealized channel functionalities implements the security properties for multi-hop payments (e.g., dishonest parties cannot steal money).
As the formalized protocol specification refines the specification of idealized channel functionalities, it follows that the formalized protocol also implements the security properties.

We describe the Lightning Network's protocol in more detail in \cref{sec-fundamentals} and give a brief summary of related work in \cref{sec-related-work}.
We give an overview of our approach in \cref{sec-proof-sketch} and describe how we show the individual proof steps.
Our work is still in-progress and, thus, not all steps are complete but with this work-in-progress report we aim to give an introduction to the overall approach.
We do not provide the proof steps in detail but elaborate on the proof ideas.

\section{Fundamentals}
\label{sec-fundamentals}

\subsection{Payment Channels: Single-Hop Payments}

A payment channel is a protocol between two users that enables these two users to deposit coins into the payment channel during opening, perform transactions between the two users by updating the payment channel, and retrieving their final funds by closing the payment channel.
At every state of the protocol, each user is guaranteed to be able to close the channel to retrieve their current balance independent of cooperation of the other user. Even with an actively malicious channel partner, an honest user cannot loose their funds as long as the user actively monitors the underlying blockchain and reacts to malicious closing attempts.

On a high level, a payment channel is implemented as a shared account: The two users open the shared account by depositing coins into the shared account and store the allocation of the funds, i.e., which user owns how many coins, in their current state.
To perform a transaction sending $x$ coins from one user to the other user, both users agree on a new allocation of funds in which $x$ coins are deducted from the sender's share of funds and added to the recipient's share of funds.
By updating their state, both users can perform an unlimited amount of transactions between each other just based on communication between each other.
To fulfill the security guarantees, it needs to be ensured that the payment channel can be closed only in a state that represents the latest allocation of funds.
Particularly, the channel may not be closed with an outdated allocation of funds and an honest user must be able to close a channel in a state with the latest allocation of funds.

More technically, a payment channel is opened in the Lightning Network by creating a funding transaction\footnote{See BOLT 2 and BOLT 3, \url{https://github.com/lightning/bolts/blob/master/00-introduction.md}.}.
The funding transaction has an input spending an output from the funding user (funder)\footnote{At present, the Lightning Network supports only single-funded channels, i.e. only one user deposits coins into the channel during opening.}
and the funding transaction has a multi-sig output that is spendable only by the two users in the channel together.
Just publishing the funding transaction on the blockchain would create a dependence of the funder on the other user for spending the funding transaction's output as the funding transaction's output can only be spent by the two users together.
To prevent such a dependence, an initial commitment transaction that spends the funding transaction's output is created by the two users and the non-funding user sends their signature for the initial commitment transaction to the funder who only publishes the funding transaction after receiving this signature.
A commitment transaction has at least two outputs: One output for each user that is redeemable only by this user and has an amount that corresponds to the balance the user currently has.
In the initial commitment transaction, all funds are spendable by the funder.\footnote{This is a simplification; the Lightning Network's specification allows the funder to send a small amount to the non-funding user already in the initial commitment transaction (see \texttt{push\_msat}).}

For a payment from one user to the other, an HTLC (Hash Timelocked Contract) is added to the channel. These HTLCs will also be used for multi-hop payments.
An HTLC is a contract that encodes the agreement that the recipient receives a specified amount if the recipient proves knowledge of a preimage to a specified hash before a specified time has passed.
To make a payment using an HTLC, the channel is updated to add the HTLC.
The HTLC is added by adding a dedicated output that represents the HTLC to a new commitment transaction. The amount of coins that are part of the HTLC are deducted from the payment's sender's output in the new commitment transaction.
After the HTLC is committed to the payment channel, the recipient of the payment fulfills the HTLC by sending the preimage to the sender of the payment.
Then, the channel is updated by creating a new commitment transaction without the HTLC output to remove the HTLC and, in the new commitment transaction, the HTLC's amount is added to the recipient's balance.
If the recipient does not fulfill the HTLC before the timelock, the HTLC is also removed but the HTLC's amount is added back to sender's balance.

For an update of the channel, the sender of the payment creates a new commitment transaction and sends a signature for this new commitment transaction to the payment's recipient.
Now, the recipient has two valid versions of the commitment transaction: The current and the new commitment transaction which are both signed by the payment's sender.
Both versions of the commitment transaction are valid and can be published on the blockchain.
As a malicious user might publish an outdated commitment transaction, commitment transactions should be `revoked' so that they cannot be published anymore.
As a signature to a commitment transaction cannot be undone, the Lightning Network uses an approach for revocation that relies on incentives: A user can be punished for publishing an outdated commitment transaction.
For each commitment transaction there exists a revocation key pair.
With knowledge of the private revocation key, one user can spend all outputs of the commitment transaction that the user's counterpart in the channel has published.
In this way, the transaction's outcome is revoked while the transaction itself is persisted.
During an update of a channel, both users send each other their signature for the new commitment transaction and reply by sending the private revocation key for the now outdated commitment transaction to revoke the outdated commitment transaction.
As the users do not have the private revocation key for the current commitment transaction of their counterpart, they cannot punish each other for correct behavior like publishing the current commitment transaction.
For the security of the protocol it is crucial that each user has the necessary private revocation keys for the states that are outdated and that the other user in the channel does not have the private revocation key for a state that is considered the latest state.

\subsection{Payment Channel Networks: Multi-Hop Payments}

If two users do not have a common payment channel but they are connected over a path of payment channels of other users, they can make multi-hop payments between each other.
The intermediate users forward the payment over their channels and might receive a small fee for their service.
To prevent intermediaries from stealing or loosing coins, it should be guaranteed for a multi-hop payment that each intermediary receives an incoming payment on one channel iff the intermediary forwards the payment on another channel.
Also the sender should send the payment to an intermediary iff the recipient receives the payment from an intermediary.
The Lightning Network uses HTLCs for multi-hop payments to achieve these security properties.
The recipient of a payment draws a random value $x$ and calculates the hash value $y = H(x)$ using a cryptographic hash function $H$.
The recipient sends $y$ to the sender of the payment.
The sender of the payment creates an HTLC with the first intermediary using $y$ as the hash condition for the HTLC.
The intermediary creates an HTLC with the next hop and each intermediary repeats this process until the last intermediary creates an HTLC with the recipient of the payment.
The recipient knows the preimage $x$ for the hash condition $y$ and fulfills the HTLC by sending $x$ to the last intermediary.
By fulfilling the HTLC, the payment's recipient receives the payment's amount from the last intermediary.
Again, each intermediary forwards the secret value $x$ fulfilling the HTLCs along the route until the sender receives $x$ and pays the first intermediary.
The timelocks of the HTLCs are chosen in a descending order from the sender to the recipient, so that each intermediary has enough time to fulfill the incoming HTLC from the previous hop if the next hop fulfills the outgoing HTLC.

\subsection{TLA\textsuperscript{+}}

The Temporal Logic of Actions (TLA) \cite{lamport_temporal_1994} is a temporal logic to reason about properties of a system. The language TLA\textsuperscript{+} is based on TLA and can be used to formalize the behavior of system.
Using tools like a model checker (TLC) or a theorem prover (TLAPS), invariants and properties can be shown to be valid for a formalized system.
In TLA\textsuperscript{+}, the state of a system is described by a set of variables $vars$.
A system is defined by defining a set of initial states for which the formula $Init$ is valid and by defining an action $Next$ that determines which steps are allowed for the system to change its state.
Using these components, a system is represented as a formula $Init \land \Box [ Next ]_{vars}$.
An additional conjunct may be a fairness condition that asserts that certain steps must be taken if they are continuously allowed.
The $Next$ action is typically a disjunct of multiple subactions the define different ways for the system's state to be updated.
These (sub)actions can be grouped into modules.
Each module can be instantiated multiple times for different sets of variables.

\section{Related Work}
\label{sec-related-work}

TLA\textsuperscript{+} is used in the industry \cite{newcombe_how_2015,resch_using_2017} and there are also examples in the scientific literature how TLA\textsuperscript{+} has been used to reason about the properties of protocols:
Narayana et al. \cite{narayana_automatic_2006} used TLA\textsuperscript{+} to search for vulnerabilities in IEEE 802.16 WiMAX protocols.
Lu et al. \cite{lu_towards_2011,lu_formal_2015} used TLA\textsuperscript{+} to verify properties of core algorithms of the Pastry protocol.
Braithwaite et al. \cite{braithwaite_formal_2020} used TLA\textsuperscript{+} for  specifying and model checking a core protocol of Tendermint blockchains.
Further, TLA\textsuperscript{+} was used to verify firewalls \cite{kim_formal_2020}, the ZooKeeper atomic broadcast protocol \cite{yin_specification_2020}, a design for state channels \cite{close_breaking_2020}, for checking security properties of smart contracts \cite{kolb_quartz_2020}, and for proving properties of Cross-Chain swaps \cite{nehai_tla_2022-1}.

The Lightning Network's protocol was formalized before by Kiayias and Thyfronitis Litos \cite{kiayias_composable_2020}.
They formalized an ideal functionality and used the UC framework \cite{canetti_universally_2001} to prove that the Lightning Network's protocol securely implements this ideal functionality.
Compared to our formalization, the protocol formalization of \cite{kiayias_composable_2020} considers more details about the cryptographic aspects.
While working on our TLA\textsuperscript{+} formalization of the Lightning Network's protocol, we found two subtle flaws in the formalization of \cite{kiayias_composable_2020} of the Lightning Network's protocol that render the formalized protocol insecure.
However, we believe that these flaws can be corrected and that the Lightning Network's protocol fulfills the ideal functionality formalized in \cite{kiayias_composable_2020}.
The first flaw concerns an incomplete description of how a user reacts to maliciously published outdated transactions.
The second flaw is more subtle and concerns how the data in an input is linked to the spending methods of an output that is spent by this input.
A detailed description of the flaws can be found in \cref{sec-appendix-composable}.
While we found the first flaw by comparison of our formalization to the formalization in the paper, we found the second flaw only by model checking when we had a similar flaw in a draft of our formalization.
While the specific flaws can be fixed with low effort, it is difficult and tedious to find such flaws in a pen-and-paper proof.
Using our approach of model checking instead, such issues can be revealed automatically.

\section{Verification of Security Properties of the Lightning Network's Protocol}
\label{sec-proof-sketch}

\subsection{Overview}

\subsubsection{Formalization of the Lightning Network's Protocol}
For the formalization of the specification of the protocol we build upon and extend the work of Grundmann et al. \cite{grundmann_verifying_2022}.
The formalization describes all possible actions how a user of the payment channel initiates transactions or reacts to messages or events.
Messages are exchanged by the two users inside a payment channel by writing messages to a message queue per user. Messages can be arbitrarily delayed but are delivered in-order.
In its structure, the formalization of the protocol specification follows the specification of the Lightning Network\footnote{\url{https://github.com/lightning/bolts/blob/master/02-peer-protocol.md}}.
The formalization abstracts, however, multiple implementation details and parts that are not part of the main functionality such as fees and error messages.
The TLA\textsuperscript{+} specification of the protocol consists of three modules:
Two modules concern the specification of actions that a user performs for the execution of the payment channel protocol:
HTLCUser specifies the actions concerning HTLCs for multi-hop payments, e.g., sending an invoice, creating an HTLC, fulfilling an HTLC.
PaymentChannelUser specifies how the payment channel is created, how the payment channel is updated when a new HTLC is added or a fulfilled HTLC is persisted, how the payment channel is closed, how an adversary can cheat, and how the honest user punishes a cheating user.
More specifically, these actions include for example actions for creating and sending a signature of a new commitment transaction to the other user, processing messages from the other user, or publishing a commitment transaction on the blockchain to close the channel.
The third module is LedgerTime, the clock that increases the current time. Time is measured in the Lightning Network's protocol by the block count of the Bitcoin blockchain. Thus, it is represented as a natural number and increased in integer steps.
The specification puts these three modules together by having a single LedgerTime module and by instantiating the HTLCUser module for each modeled user and one instance of PaymentChannelUser per channel for each user.
Formally, the TLA\textsuperscript{+} specification is defined by a set of initial states and a $Next$ action that describes possible steps that can lead from one state to a new state.
The $Next$ action for a specification with three users and two channels can be found in \cref{fig-tla-full-spec}.
This formal specification is pictorially represented in \cref{fig-img-full-spec}.

\begin{figure}
\begin{tlatex}
\@x{ Next \.{\defeq}}%
\@x{\@s{16.4} \.{\lor} LedgerTime}%
\@x{\@s{16.4} \.{\lor} HTLCUser(Alice)}%
\@x{\@s{16.4} \.{\lor} HTLCUser(Bob)}%
\@x{\@s{16.4} \.{\lor} HTLCUser(Charlie)}%
\@x{\@s{16.4} \.{\lor} PaymentChannelUser(AB, Alice)}%
\@x{\@s{16.4} \.{\lor} PaymentChannelUser(AB, Bob)}%
\@x{\@s{16.4} \.{\lor} PaymentChannelUser(BC, Bob)}%
\@x{\@s{16.4} \.{\lor} PaymentChannelUser(BC, Charlie)}%
\end{tlatex}
\caption{$Next$ action of the specification of a payment channel network with three users and two payment channels.}
\label{fig-tla-full-spec}
\end{figure}

\begin{figure}
\includegraphics[width=\linewidth]{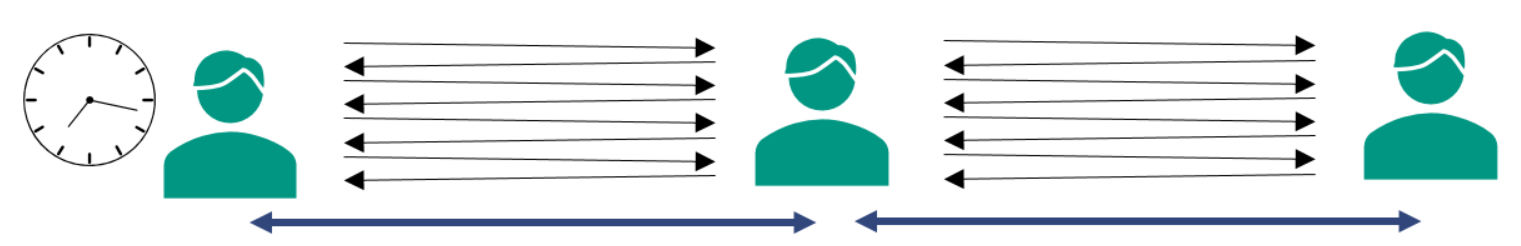}
\caption{Pictorial representation of the full specification of a payment channel network with three users and two payment channels. The clock represents the LedgerTime module; the blue lines represent three users who exchange messages through the HTLCUser module; the black arrows represent a payment channel in which two instances of the PaymentChannelUser module exchange messages.}
\label{fig-img-full-spec}
\end{figure}

\subsubsection{Security Properties}
Our goal is to show that this specification implements functional properties and fulfills security properties, e.g., (1) an honest user finally receives the user's correct balance even if other users act maliciously by publishing outdated states on the blockchain, or (2) the balances after a multi-hop payment are correct, i.e., the payment's amount is deducted from the sender's balance, any intermediate's balance stays the same\footnote{An intermediate's balances change in the two payment channels that the intermediate uses for forwarding a payment but the overall balance of the intermediate in all payment channels of the intermediate should stay the same.}, and the payment's amount is added to the recipient's balance.
We formalized the security properties in form of an idealized payment network functionality whose full specification can be found in \cref{sec-appendix-idealized-payment-network}.
This ideal payment network functionality models a payment network in which users start with an external balance (stored in variable ExtBalances), deposit assets, make payments (variable Payments), and withdraw their assets.
User can be dishonest (variable Honest). The ideal functionality specifies that dishonest users cannot steal money from other users. Instead, dishonest users might be punished for cheating by loosing a part of their balance.
A system for which the external variables ExtBalances, Payments, and Honest have values that are allowed by the ideal payment network functionality is secure, i.e., balances are computed correctly and dishonest users can only loose but not gain assets.

\subsubsection{Challenge: Exploring the State Space}
Because the order of how messages are sent and processed in the payment channel protocol can vary, there are many different possible executions of the protocol.
The state space explodes if two or more payment channels are modeled because there is a large amount of different combinations of the states the payment channels can be in.
Therefore, a specification modeling multiple payment channels is too large for model checking.
To verify the security properties of a specification for multiple payment channels nevertheless, we use the following approach.

\subsubsection{Approach: Abstracting Specification of a Payment Channel }
We specify an idealized multi-hop specification that uses an ideal functionality of a payment channel instead of a formalization of the real protocol.
For the idealized multi-hop specification, we replace the two instances of PaymentChannelUsers per channel by an instance of the module IdealChannel. A pictorial representation is shown in \cref{fig-img-pcn-with-ideal} and the $Next$ action is shown in \cref{fig-tla-pcn-with-ideal}.
The module IdealChannel specifies the functionality of a payment channel on a coarser granularity: The actions describe the changes to both parties' state simultaneously and abstract from the exchange of messages on the protocol level as specified in PaymentChannelUser.
Abstracting the behavior of one payment channel reduces the state space and, together with an optimization of the LedgerTime module, it allows for model checking the combination of multiple payment channels in multi-hop payments.

\begin{figure}
\begin{tlatex}
\@pvspace{8.0pt}%
\@x{ Next \.{\defeq}}%
\@x{\@s{16.4} \.{\lor} LedgerTime}%
\@x{\@s{16.4} \.{\lor} HTLCUser(Alice)}%
\@x{\@s{16.4} \.{\lor} HTLCUser(Bob)}%
\@x{\@s{16.4} \.{\lor} HTLCUser(Charlie)}%
\@x{\@s{16.4} \.{\lor} IdealChannel(AB)}%
\@x{\@s{16.4} \.{\lor} IdealChannel(BC)}%
\@pvspace{8.0pt}%
\end{tlatex}
\caption{$Next$ action of a payment channel network using the ideal functionality for payment channels. }
\label{fig-tla-pcn-with-ideal}
\end{figure}

\begin{figure}
\includegraphics[width=\linewidth]{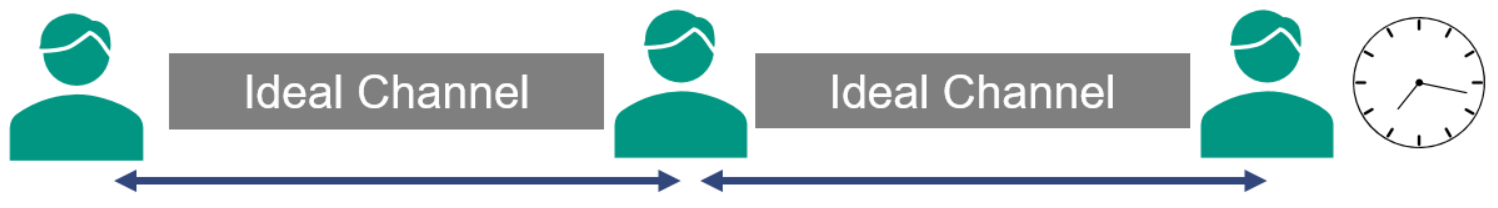}
\caption{Pictorial representation of the idealized multi-hop specification of a payment channel network with three users and two payment channels. Each ideal channel is modeled by an instance of IdealChannel.}
\label{fig-img-pcn-with-ideal}
\end{figure}

We use the idealized multi-hop specification to show that the specification of the Lightning Network's protocol fulfills the security properties:
We show that the idealized multi-hop specification fulfills the security properties and we extend this result to the protocol specification by showing that the protocol specification is a refinement of the idealized multi-hop specification.
The protocol specification refines the idealized multi-hop specification iff for every behavior of the protocol specification there exists a behavior of the idealized multi-hop specification for which the externally visible variables (i.e., ExtBalances, Payments, Honest) have the same values.
As the security properties rely on the externally visible variables only, the protocol specification fulfills the security properties if the idealized multi-hop specification does.
From a security perspective the refinement means that every attack that is possible in the protocol specification must also be possible in the idealized multi-hop specification.
Showing the absence of attacks in the idealized multi-hop specification, thus, shows that the protocol specification is secure.

\begin{figure*}
\includegraphics[width=\linewidth]{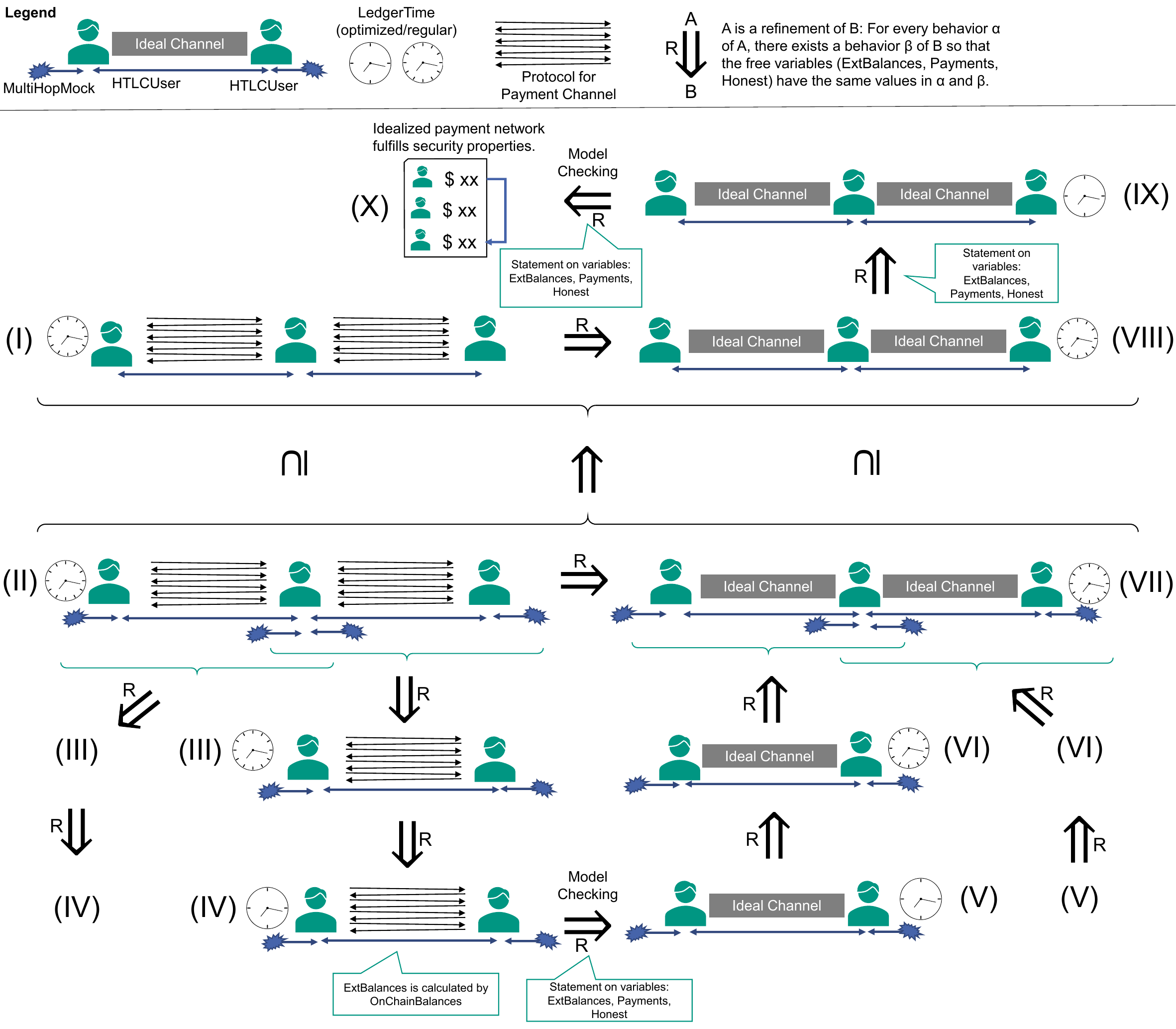}
\caption{Proof sketch showing how we show that the protocol for multi-hop payments (I) implements the idealized payment network (X) which fulfills the security properties.
The definitions of the $Next$ actions of specifications $(I)$ to $(IX)$ can be found in \cref{sec-appendix-next-specs}. The definition of specification $(X)$ is shown in \cref{sec-appendix-idealized-payment-network}.
}
\label{fig-img-proof-sketch}
\end{figure*}

\subsection{Proof sketch}

In the following, we given an overview of our proof and the most important arguments.
The proof's structure is graphically shown in \cref{fig-img-proof-sketch}.
In the following, we refer to specifications with Roman numbers $(I)$ to $(X)$ as represented in \cref{fig-img-proof-sketch} and (partially) defined in \cref{sec-appendix-next-specs}.
In the proof sketch shown in in \cref{fig-img-proof-sketch}, the ideal payment network functionality is depicted as $(X)$.
Our goal is to show that the specification of the Lightning Network's protocol $(I)$ fulfills the security properties, i.e. we show the validity of the formula $(I) \Rightarrow (X)$ in which the free variables are ExtBalance, Payments, and Honest.

As specification $(I)$ is too complex for model checking, we specify two abstractions of this specification ($(VIII)$ and $(IX)$) so that abstraction $(IX)$ can be model checked.
The first abstraction step is to abstract the formalization of multi-hop payments using the Lightning Network's protocol to a formalization of multi-hop payments using the idealized channel specification ($(I) \Rightarrow (VIII)$).
The second abstraction is to abstract time by grouping equivalent behaviors that differ only by the timestamps ($(VIII) \Rightarrow (IX)$).
Then, we use a model checker to show that multi-hop payments with idealized payment channels $(IX)$ are a refinement of the idealized payment network specification $(X)$.

To show that $(I) \Rightarrow (VIII)$, we decompose the specification of multiple payment channels into the specification of a single payment channel and show that each payment channel implemented by the Lightning Network's protocol is a refinement of the idealized payment channel specification.
To include a specification of the environment of a single payment channel, we specify the module MultiHopMock. An instance of the module MultiHopMock for one specific user abstracts the behavior that other payment channels and users can have on this user's variables, i.e., every action that can happen in the payment channel between Alice and Bob is an action of the instance of MultiHopMock for Charlie who has a channel with Bob.
Having the module MultiHopMock abstracting other users and payment channels allows us to compose the specifications of single payment channels back to multi-hop payments when using the idealized specification for a payment channel ($(VI) \Rightarrow (VII)$).
We show that multi-hop payments using the Lightning Network's protocol and the MultiHopMock are a refinement of multi-hop payments using the idealized channel specification and the MultiHopMock ($(II) \Rightarrow (VII)$).

We show that $(II) \Rightarrow (VII)$ by first decomposing the specification of multiple payment channels into a single payment channel ($(II) \Rightarrow (IV)$).
Then, we show by specifying a refinement mapping verified by a model checker that the specification of the Lightning Network's protocol refines the idealized channel specification ($(IV) \Rightarrow (V)$).
As this step reduces the complexity of the Lightning Network's protocol into a simpler ideal functionality, this step is the most difficult part of the proof. To verify the correctness of the step, we use a model checker.
Then, we show how this result extends to multi-hop payments using idealized payment channels ($(V) \Rightarrow (VII)$).
Having shown that there exists a refinement mapping for $(II) \Rightarrow (VII)$, we show that $(I) \Rightarrow (VIII)$ by showing that the subset of specification $(II)$ that equals specification $(I)$ is mapped by the refinement mapping to the subset of specification $(VII)$ that equals specification $(VIII)$.

In the following, we elaborate on the intuition behind these proof steps.

\subsubsection{$(II) \Rightarrow (III)$}

Compared to the protocol specification $(I)$, the specification $(II)$ adds for each modeled user an instance of MultiHopMock, a module that abstracts effects that other users can have on one user.
We show that each single payment channel in specification $(II)$ behaves as specified by specification $(III)$.
In informal words, this means that if we look at only the variables that are relevant for one single payment channel then specification $(II)$ refines specification $(III)$.
Specification $(II)$ consists of an instance of LedgerTime, one instance of HTLCUser per user, one instance of MultiHopMock per modeled user, and two instances of PaymentChannelUser per payment channel.
A step of the specification $(II)$ can be a step of any of these modules.
We refer to the set of variables of specification $(II)$ that concern the payment channel $AB$ as $v_{AB}$.
Considering $v_{AB}$, we show that the instances of PaymentChannelUser of any other payment channel and the instances of HTLCUser of any user not being part of the payment channel $AB$ are a refinement of the instance of MultiHopMock for channel $AB$.
Intuitively, this means for a channel $AB$ that every step that happens in another payment channel does either not affect the variables of $AB$ or is also a step by MultiHopMock which is also part of specification $(III)$.
Therefore, each step of specification $(II)$ that changes the variables $v_{AB}$, is also a step of specification $(III)$ for the same set of variables.

\subsubsection{$(III) \Rightarrow (IV)$}

While specification $(III)$ includes an instance of LedgerTime that allows incrementing the current time by single time steps, specification $(IV)$ uses an instance of an optimized LedgerTime module that skips points in time that lead to equivalent future behaviors.
The intuition why this is possible is that, in the payment channel protocol, the current time is only used for checking whether a point in time has already passed or not.
For each such condition, two points in time that are both on the same side of the comparison lead to the same possible steps.
We show that $(III) \Rightarrow (IV)$ by showing that for each behavior of specification $(III)$ there exists a behavior of specification $(IV)$ where the time might be different but all other variables have the same values.

\subsubsection{$(IV) \Rightarrow (V)$}

On the one hand, showing that specification $(IV)$ formalizing the Lightning Network's protocol is a refinement of specification $(V)$ formalizing an idealized payment channel is the most difficult refinement of the chain of refinements that we show.
On the other hand, the simplifications of the previous refinements from specification $(I)$ to $(IV)$ result in specification $(IV)$ having a state space that is explorable using model checking.
We show that $(IV) \Rightarrow (V)$ by explicitly formalizing a refinement mapping between specification $(IV)$ and specification $(V)$ and validating this refinement mapping by model checking.

\subsubsection{$(V) \Rightarrow (VI)$}

Specification $(V)$ trivially refines specification $(VI)$ because the only change between the two specifications is that specification $(V)$ uses an optimized LedgerTime instead of the regular LedgerTime and each step of the optimized LedgerTime is also a step of the regular LedgerTime module.

\subsubsection{$(VI) \Rightarrow (VII)$}

Compared to specification $(VI)$ for a single payment channel, specification $(VII)$ composes multiple payment channels.
A behavior of specification $(VI)$ is also a behavior of specification $(VII)$ in which no steps are taken in other payment channels.
Thus, $(VI) \Rightarrow (VII)$.

\subsubsection{$(I) \Rightarrow (VIII)$}

From above, we conclude that $(II) \Rightarrow (VII)$.
The composition of refinement mappings of the individual steps defines a refinement mapping from $(II)$ to $(VII)$ that we call $f$ in the following.
We argue that the protocol implementation without the module MultiHopMock also refines the idealized specification without the module MultiHopMock ($(I) \Rightarrow (VIII)$) because restricting the domain of the refinement mapping $f$ to the behaviors allowed by specification $(I)$ results in a refinement mapping from $(I)$ to $(VII)$.
This follows from the fact that by construction of the individual refinement mappings that the refinement mapping $f$ is composed of, the refinement mapping $f$ affects only steps of the module PaymentChannelUser which are mapped to stuttering steps or steps of the module IdealChannel.
The refinement mapping $f$ maps a step of the module MultiHopMock or the module HTLCUser or the module LedgerTime in specification $(II)$ to a step of MultiHopMock (resp. HTLCUser, LedgerTime) in specification $(VII)$.
It follows that the refinement mapping $f$ is a also a refinement mapping for $(I) \Rightarrow (VIII)$.

\subsubsection{$(VIII) \Rightarrow (IX)$}

To get a specification with a reduced state space, we replace the regular LedgerTime module in specification $(VIII)$ by an optimized LedgerTime module and retrieve specification $(IX)$.
Analogously to the refinement $(III) \Rightarrow (IV)$, this optimization does not drop any behaviors and, thus, $(VIII) \Rightarrow (IX)$.

\subsubsection{$(IX) \Rightarrow (X)$}

Through model checking, we show that specification $(IX)$ is a refinement of the idealized payment network functionality $(X)$ and, thus, fulfills the security properties.

\subsubsection{Conclusion}

From the refinements above, we conclude that $(I) \Rightarrow (X)$, i.e. the specification of the Lightning Network's protocol is an implementation of an idealized payment network and fulfills the security properties.

\section{Conclusion}

While the Lightning Network's protocol is complex and many different states can be reached by a model checker, we have presented an approach that makes it possible to verify the security of the protocol in a partially machine-checked way.
The approach uses model checking for two proof steps that are difficult to reason about: Showing that a single payment channel refines the ideal channel functionality and showing that the protocol for multi-hop payments is secure.
We are currently working on fully formalizing every step of the proof.

\bibliographystyle{IEEEtran}
\bibliography{library}

\appendix

\subsection{Idealized Payment Network Specification}
\label{sec-appendix-idealized-payment-network}

\cref{fig-tla-idealized-payment-network} shows the specification of an idealized payment network.
It can easily be checked that this specification is secure, i.e., an honest party can withdraw at least their correct balance.
The $Next$ action of the specification is a disjunct of five actions:
DepositBalance is for one user to deposit balance into the payment network.
WithdrawBalance can be used to withdraw the balance of one or two users from the payment network.
ProcessPayment processes a payment that has initially been specified in the Payments variable. This action simply removes the payment's amount from the sender's balance and adds the same amount to the recipient's balance.
AbortPayment models aborted payments by removing the payment from the Payment variable without any further change to the state.
PunishCheating models a cheating dishonest party that is punished by another party. 
As a punishment the punishing party retrieves a part of the dishonest party's balance.

\begin{figure*}
\begin{tlatex}
\@xx{}%
\@pvspace{8.0pt}%
 \@x{}\moduleLeftDash\@xx{ {\MODULE}
 IdealPaymentNetwork}\moduleRightDash\@xx{}%
\@pvspace{8.0pt}%
\@x{ {\EXTENDS} Naturals ,\, Sequences ,\, FiniteSets}%
\@x{ {\VARIABLES} ExternalBalances ,\, Payments ,\, Honest ,\, Balances}%
 \@x{ vars \.{\defeq} {\langle} ExternalBalances ,\, Balances ,\, Payments ,\,
 Honest {\rangle}}%
\@x{ {\CONSTANT} NumUsers}%
\@x{ UserIds \.{\defeq} 1 \.{\dotdot} NumUsers}%
\@pvspace{8.0pt}%
\@x{ Init \.{\defeq}}%
 \@x{\@s{16.4} \.{\land} ExternalBalances \.{\in} [ UserIds \.{\rightarrow}
 Nat ]}%
\@x{\@s{16.4} \.{\land} Balances \.{=} [ i \.{\in} UserIds \.{\mapsto} 0 ]}%
 \@x{\@s{16.4} \.{\land} Payments \.{\in} {\SUBSET} [ amount \.{\mapsto} Nat
 ,\, sender \.{\mapsto} UserIds ,\, recipient \.{\mapsto} UserIds ]}%
 \@x{\@s{16.4} \.{\land} Honest \.{\in} [ UserIds \.{:} \{ {\TRUE} ,\,
 {\FALSE} \} ]}%
\@pvspace{8.0pt}%
\@x{ DepositBalance \.{\defeq}}%
\@x{\@s{16.4} \.{\land} \E\, user \.{\in} UserIds \.{:}}%
 \@x{\@s{20.5} \.{\land} \E\, amount \.{\in} 1 \.{\dotdot} ExternalBalances [
 user ] \.{:}}%
 \@x{\@s{24.6} \.{\land} ExternalBalances \.{'} \.{=} [ ExternalBalances
 {\EXCEPT} {\bang} [ user ] \.{=} @ \.{-} amount ]}%
 \@x{\@s{24.6} \.{\land} Balances \.{'} \.{=} [ Balances {\EXCEPT} {\bang} [
 user ] \.{=} @ \.{+} amount ]}%
\@x{\@s{16.4} \.{\land} {\UNCHANGED} {\langle} Payments ,\, Honest {\rangle}}%
\@pvspace{8.0pt}%
\@x{ WithdrawBalance \.{\defeq}}%
 \@x{\@s{16.4} \.{\land} \E\, userA \.{\in} UserIds \.{:} \E\, userB \.{\in}
 UserIds \.{:}}%
 \@x{\@s{20.5} \.{\land} \E\, amountA \.{\in} 1 \.{\dotdot} Balances [ userA ]
 \.{:} \E\, amountB \.{\in} 0 \.{\dotdot} Balances [ userB ] \.{:}}%
 \@x{\@s{24.6} \.{\land} Balances \.{'} \.{=} [ Balances {\EXCEPT} {\bang} [
 userA ] \.{=} @ \.{-} amountA ,\, {\bang} [ userB ] \.{=} @ \.{-} amountB ]}%
 \@x{\@s{24.6} \.{\land} ExternalBalances \.{'} \.{=} [ ExternalBalances
 {\EXCEPT} {\bang} [ userA ] \.{=} @ \.{+} amountA ,\, {\bang} [ userB ]
 \.{=} @ \.{+} amountB ]}%
\@x{\@s{16.4} \.{\land} {\UNCHANGED} {\langle} Payments ,\, Honest {\rangle}}%
\@pvspace{8.0pt}%
\@x{ ProcessPayment \.{\defeq}}%
\@x{\@s{16.4} \.{\land} \E\, payment \.{\in} Payments \.{:}}%
 \@x{\@s{24.59} \.{\land} Balances \.{'} \.{=} [ Balances {\EXCEPT} {\bang} [
 payment . sender ] \.{=} @ \.{-} payment . amount ,\, {\bang} [ payment .
 recipient ] \.{=} @ \.{+} payment . amount ]}%
 \@x{\@s{24.59} \.{\land} Payments \.{'} \.{=} Payments \.{\,\backslash\,} \{
 payment \}}%
 \@x{\@s{16.4} \.{\land} {\UNCHANGED} {\langle} ExternalBalances ,\, Honest
 {\rangle}}%
\@pvspace{8.0pt}%
\@x{ AbortPayment \.{\defeq}}%
 \@x{\@s{16.4} \.{\land} \E\, payment \.{\in} Payments \.{:} Payments \.{'}
 \.{=} Payments \.{\,\backslash\,} \{ payment \}}%
 \@x{\@s{16.4} \.{\land} {\UNCHANGED} {\langle} ExternalBalances ,\, Balances
 ,\, Honest {\rangle}}%
\@pvspace{8.0pt}%
\@x{ PunishCheating \.{\defeq}}%
\@x{\@s{16.4} \.{\land} \E\, dishonestUser \.{\in} UserIds \.{:}}%
\@x{\@s{20.5} \.{\land} Honest [ dishonestUser ] \.{=} {\FALSE}}%
\@x{\@s{20.5} \.{\land} \E\, otherUser \.{\in} UserIds \.{:}}%
\@x{\@s{24.6} \.{\land} otherUser \.{\neq} dishonestUser}%
 \@x{\@s{24.6} \.{\land} \E\, amount \.{\in} 1 \.{\dotdot} Balances [
 dishonestUser ] \.{:}}%
 \@x{\@s{28.7} \.{\land} Balances \.{'} \.{=} [ Balances {\EXCEPT} {\bang} [
 dishonestUser ] \.{=} @ \.{-} amount ,\, {\bang} [ otherUser ] \.{=} @ \.{+}
 amount ]}%
 \@x{\@s{16.4} \.{\land} {\UNCHANGED} {\langle} ExternalBalances ,\, Honest
 {\rangle}}%
\@pvspace{8.0pt}%
\@x{ Next \.{\defeq}}%
\@x{\@s{16.4} \.{\lor} DepositBalance}%
\@x{\@s{16.4} \.{\lor} ProcessPayment}%
\@x{\@s{16.4} \.{\lor} AbortPayment}%
\@x{\@s{16.4} \.{\lor} PunishCheating}%
\@x{\@s{16.4} \.{\lor} WithdrawBalance}%
\@pvspace{8.0pt}%
\@x{ Spec \.{\defeq}}%
\@x{\@s{16.4} \.{\land} Init}%
\@x{\@s{16.4} \.{\land} {\Box} [ Next ]_{ vars}}%
\@x{\@s{16.4} \.{\land} {\WF}_{ vars} ( WithdrawBalance )}%
\@pvspace{8.0pt}%
\@x{}\bottombar\@xx{}%
\end{tlatex}
    \caption{Idealized payment network that fulfills the security properties.}
    \label{fig-tla-idealized-payment-network}
\end{figure*}

\subsection{Next Actions of Specifications}
\label{sec-appendix-next-specs}

Each specification shown in \cref{fig-img-proof-sketch}, is defined by an $Init$ predicate that defines the set of initial states, a $Next$ action that defines possible steps, and a fairness condition.
In the following, we show how the $Next$ action of each specification is composed.
The $Next$ actions of the modules used in these definitions are shown in \cref{sec-appendix-next-modules}.

\begin{tlatex}
\@pvspace{8.0pt}%
\@x{ Next\_I \.{\defeq}}%
\@x{\@s{16.4} \.{\lor} LedgerTime\_Next}%
\@x{\@s{16.4} \.{\lor} HTLCUserNext ( Alice )}%
\@x{\@s{16.4} \.{\lor} HTLCUserNext ( Bob )}%
\@x{\@s{16.4} \.{\lor} HTLCUserNext ( Charlie )}%
\@x{\@s{16.4} \.{\lor} PaymentChannelUserNext ( AB ,\, Alice )}%
\@x{\@s{16.4} \.{\lor} PaymentChannelUserNext ( AB ,\, Bob )}%
\@x{\@s{16.4} \.{\lor} PaymentChannelUserNext ( BC ,\, Bob )}%
\@x{\@s{16.4} \.{\lor} PaymentChannelUserNext ( BC ,\, Charlie )}%
\@pvspace{8.0pt}%
\@x{ Next\_II \.{\defeq}}%
\@x{\@s{16.4} \.{\lor} LedgerTime\_Next}%
\@x{\@s{16.4} \.{\lor} HTLCUserNext ( Alice )}%
\@x{\@s{16.4} \.{\lor} HTLCUserNext ( Bob )}%
\@x{\@s{16.4} \.{\lor} HTLCUserNext ( Charlie )}%
\@x{\@s{16.4} \.{\lor} PaymentChannelUserNext ( AB ,\, Alice )}%
\@x{\@s{16.4} \.{\lor} PaymentChannelUserNext ( AB ,\, Bob )}%
\@x{\@s{16.4} \.{\lor} PaymentChannelUserNext ( BC ,\, Bob )}%
\@x{\@s{16.4} \.{\lor} PaymentChannelUserNext ( BC ,\, Charlie )}%
\@x{\@s{16.4} \.{\lor} MultiHopMockNext ( Alice )}%
\@x{\@s{16.4} \.{\lor} MultiHopMockNext ( Bob )}%
\@x{\@s{16.4} \.{\lor} MultiHopMockNext ( Charlie )}%
\@pvspace{8.0pt}%
\@x{ Next\_III \.{\defeq}}%
\@x{\@s{16.4} \.{\lor} LedgerTime\_Next}%
\@x{\@s{16.4} \.{\lor} HTLCUserNext ( Alice )}%
\@x{\@s{16.4} \.{\lor} HTLCUserNext ( Bob )}%
\@x{\@s{16.4} \.{\lor} PaymentChannelUserNext ( AB ,\, Alice )}%
\@x{\@s{16.4} \.{\lor} PaymentChannelUserNext ( AB ,\, Bob )}%
\@x{\@s{16.4} \.{\lor} MultiHopMock\_Next}%
\@pvspace{8.0pt}%
\@x{ Next\_IV \.{\defeq}}%
\@x{\@s{16.4} \.{\lor} OptimizedLedgerTime\_Next}%
\@x{\@s{16.4} \.{\lor} HTLCUserNext ( Alice )}%
\@x{\@s{16.4} \.{\lor} HTLCUserNext ( Bob )}%
\@x{\@s{16.4} \.{\lor} PaymentChannelUserNext ( AB ,\, Alice )}%
\@x{\@s{16.4} \.{\lor} PaymentChannelUserNext ( AB ,\, Bob )}%
\@x{\@s{16.4} \.{\lor} MultiHopMock\_Next}%
\@pvspace{8.0pt}%
\@x{ Next\_V \.{\defeq}}%
\@x{\@s{16.4} \.{\lor} OptimizedLedgerTime\_Next}%
\@x{\@s{16.4} \.{\lor} HTLCUserNext ( Alice )}%
\@x{\@s{16.4} \.{\lor} HTLCUserNext ( Bob )}%
\@x{\@s{16.4} \.{\lor} IdealChannelNext ( AB )}%
\@x{\@s{16.4} \.{\lor} MultiHopMock\_Next}%
\@pvspace{8.0pt}%
\@x{ Next\_VI \.{\defeq}}%
\@x{\@s{16.4} \.{\lor} LedgerTime\_Next}%
\@x{\@s{16.4} \.{\lor} HTLCUserNext ( Alice )}%
\@x{\@s{16.4} \.{\lor} HTLCUserNext ( Bob )}%
\@x{\@s{16.4} \.{\lor} IdealChannelNext ( AB )}%
\@x{\@s{16.4} \.{\lor} MultiHopMock\_Next}%
\@pvspace{8.0pt}%
\@x{ Next\_VII \.{\defeq}}%
\@x{\@s{16.4} \.{\lor} LedgerTime\_Next}%
\@x{\@s{16.4} \.{\lor} HTLCUserNext ( Alice )}%
\@x{\@s{16.4} \.{\lor} HTLCUserNext ( Bob )}%
\@x{\@s{16.4} \.{\lor} HTLCUserNext ( Charlie )}%
\@x{\@s{16.4} \.{\lor} IdealChannelNext ( AB )}%
\@x{\@s{16.4} \.{\lor} IdealChannelNext ( BC )}%
\@x{\@s{16.4} \.{\lor} MultiHopMockNext ( Alice )}%
\@x{\@s{16.4} \.{\lor} MultiHopMockNext ( Bob )}%
\@x{\@s{16.4} \.{\lor} MultiHopMockNext ( Charlie )}%
\@pvspace{8.0pt}%
\@x{ Next\_VIII \.{\defeq}}%
\@x{\@s{16.4} \.{\lor} LedgerTime\_Next}%
\@x{\@s{16.4} \.{\lor} HTLCUserNext ( Alice )}%
\@x{\@s{16.4} \.{\lor} HTLCUserNext ( Bob )}%
\@x{\@s{16.4} \.{\lor} HTLCUserNext ( Charlie )}%
\@x{\@s{16.4} \.{\lor} IdealChannelNext ( AB )}%
\@x{\@s{16.4} \.{\lor} IdealChannelNext ( BC )}%
\@pvspace{8.0pt}%
\@x{ Next\_IX \.{\defeq}}%
\@x{\@s{16.4} \.{\lor} OptimizedLedgerTime\_Next}%
\@x{\@s{16.4} \.{\lor} HTLCUserNext ( Alice )}%
\@x{\@s{16.4} \.{\lor} HTLCUserNext ( Bob )}%
\@x{\@s{16.4} \.{\lor} HTLCUserNext ( Charlie )}%
\@x{\@s{16.4} \.{\lor} IdealChannelNext ( AB )}%
\@x{\@s{16.4} \.{\lor} IdealChannelNext ( BC )}%
\end{tlatex}

\subsection{Next Actions of Modules}
\label{sec-appendix-next-modules}

In the following, we list $Next$ actions of the modules used in our specifications.
Each action is a conjunct of multiple subactions of which the definitions are not printed.

\begin{tlatex}
\@pvspace{8.0pt}%
\@x{ LedgerTime\_Next \.{\defeq}}%
\@x{\@s{16.4} \.{\lor} AdvanceLedgerTime}%
\@pvspace{8.0pt}%
\@x{ OptimizedLedgerTime\_Next \.{\defeq}}%
\@x{\@s{16.4} \.{\lor} OptimizedAdvanceLedgerTime}%
\@pvspace{8.0pt}%
\@x{ HTLCUser\_Next \.{\defeq}}%
\@x{\@s{16.4} \.{\lor} RequestInvoice}%
\@x{\@s{16.4} \.{\lor} GenerateAndSendPaymentHash}%
\@x{\@s{16.4} \.{\lor} ReceivePaymentHash}%
\@x{\@s{16.4} \.{\lor} AddAndSendOutgoingHTLC}%
\@x{\@s{16.4} \.{\lor} ReceiveUpdateAddHTLC}%
\@x{\@s{16.4} \.{\lor} SendHTLCPreimage}%
\@x{\@s{16.4} \.{\lor} ReceiveHTLCPreimage}%
\@x{\@s{16.4} \.{\lor} SendHTLCFail}%
\@x{\@s{16.4} \.{\lor} ReceiveHTLCFail}%
\@pvspace{8.0pt}%
\@x{ IdealChannel\_Next \.{\defeq}}%
\@x{\@s{16.4} \.{\lor} OpenPaymentChannel}%
\@x{\@s{16.4} \.{\lor} UpdatePaymentChannel}%
\@x{\@s{16.4} \.{\lor} CommitHTLCsOnChain}%
\@x{\@s{16.4} \.{\lor} FulfillHTLCsOnChain}%
\@x{\@s{16.4} \.{\lor} WillPunishLater}%
\@x{\@s{16.4} \.{\lor} ClosePaymentChannel}%
\@pvspace{8.0pt}%
\@x{ PaymentChannelUser\_Next \.{\defeq}}%
\@x{\@s{16.4} \.{\lor} SendOpenChannel}%
\@x{\@s{16.4} \.{\lor} SendAcceptChannel}%
\@x{\@s{16.4} \.{\lor} CreateFundingTransaction}%
\@x{\@s{16.4} \.{\lor} SendSignedFirstCommitTransaction}%
\@x{\@s{16.4} \.{\lor} ReplyWithFirstCommitTransaction}%
\@x{\@s{16.4} \.{\lor} ReceiveCommitTransaction}%
\@x{\@s{16.4} \.{\lor} PublishFundingTransaction}%
\@x{\@s{16.4} \.{\lor} NoteThatFundingTransactionPublished}%
\@x{\@s{16.4} \.{\lor} SendNewRevocationKey}%
\@x{\@s{16.4} \.{\lor} ReceiveNewRevocationKey}%
\@x{\@s{16.4} \.{\lor} SendSignedCommitment}%
\@x{\@s{16.4} \.{\lor} ReceiveSignedCommitment}%
\@x{\@s{16.4} \.{\lor} ReceiveSignedCommitmentDuringClosing}%
\@x{\@s{16.4} \.{\lor} RevokeAndAck}%
\@x{\@s{16.4} \.{\lor} ReceiveRevocationKey}%
\@x{\@s{16.4} \.{\lor} ReceiveRevocationKeyForTimedoutHTLC}%
\@x{\@s{16.4} \.{\lor} CloseChannel}%
\@x{\@s{16.4} \.{\lor} Cheat}%
\@x{\@s{16.4} \.{\lor} Punish}%
\@x{\@s{16.4} \.{\lor} NoteThatOtherPartyClosedHonestly}%
\@x{\@s{16.4} \.{\lor} NoteThatOtherPartyClosedButUnpunishable}%
\@x{\@s{16.4} \.{\lor} NoteThatOtherPartyClosedDishonestly}%
\@x{\@s{16.4} \.{\lor} NoteCommittedAndUncommittedAndPersistedHTLCs}%
\@x{\@s{16.4} \.{\lor} NotePunishedHTLCs}%
\@x{\@s{16.4} \.{\lor} UpdatePunishedHTLCs}%
\@x{\@s{16.4} \.{\lor} NoteAbortedHTLCs}%
\@x{\@s{16.4} \.{\lor} RedeemHTLCAfterClose}%
\@x{\@s{16.4} \.{\lor} NoteThatHTLCFulfilledOnChain}%
\@x{\@s{16.4} \.{\lor} NoteThatHTLCTimedOutOnChain}%
\@x{\@s{16.4} \.{\lor} WillPunishLater}%
\@x{\@s{16.4} \.{\lor} InitiateShutdown}%
\@x{\@s{16.4} \.{\lor} ReceiveInitiateShutdown}%
\@x{\@s{16.4} \.{\lor} IgnoreMessageDuringClosing}%
\@x{\@s{16.4} \.{\lor} NoteThatChannelClosedAndAllHTLCsRedeemed}%
\@pvspace{8.0pt}%
\@x{ MultiHopMock\_Next \.{\defeq}}%
\@x{\@s{16.4} \.{\lor} AddNewForwardedPayment}%
\@x{\@s{16.4} \.{\lor} ReceivePreimageForIncomingHTLC}%
\end{tlatex}

\subsection{On the Formalization of \cite{kiayias_composable_2020}}
\label{sec-appendix-composable}

While working on the formalization of the Lightning Network's protocol in TLA\textsuperscript{+}, we found the following two flaws in the formalization of \cite{kiayias_composable_2020}.
While these flaws render the formalized protocol insecure, they are easy to fix and it seems that the security proof could work for the corrected protocol.
The following references to figures and page numbers refer to the paper's version on ePrint \cite[version 20220217:205237]{kiayias_composable_2019-1}.

The first flaw concerns the punishment of the publication of an outdated commitment transaction for which the protocol is specified in Fig. 37, lines 21-25 (page 64).
A problem arises for example in the following situation:
Before the current time, user Alice has sent an outgoing HTLC to user Bob. The HTLC was committed and has been fulfilled. Now, the HTLC's absolute timelock has passed.
Now, Alice has an outdated commitment transaction that commits the HTLC and Alice has Bob's signature on the HTLC timeout transaction corresponding to that HTLC.
Alice is malicious and publishes this outdated commitment transaction together with the HTLC timeout transaction which is valid because the HTLC's absolute timelock has passed.
Bob runs the protocol specified in Fig. 37 and arrives at line 22.
In line 22, a revocation transaction is created whose inputs spend all outputs of the outdated commitment transaction.
In the situation described, such a revocation transaction cannot be valid because the HTLC output in the outdated commitment transaction is already spent.
Instead of an input referencing the outdated commitment transaction's HTLC output, the revocation transaction must have an input that references the output of the HTLC timeout transaction.
While the protocol as formalized in Fig. 37 is incorrect, the security proof on page 90 does not mention the case that a second-stage (timeout or success) HTLC transaction might have been published for an outdated commitment transaction and, thus, the protocol seems to be correct.
While the protocol can be corrected by adding a specification of how such cases are handled, it is hard to detect such flaws by inspecting the proof manually.

For a scenario that shows the impact of the second flaw, assume that in the payment channel between users Alice and Bob there is currently an unfulfilled HTLC for a payment from Alice to Bob.
The HTLC's absolute timelock passes and the HTLC times out.
Bob unilaterally closes the payment channel by publishing the latest commitment transaction.
The commitment transaction contains an output for the HTLC with the spending method $pt_{\mathrm{rev},n+1} \lor (pt_{\mathrm{htlc}, n+1}, \mathrm{\mathtt{CltvExpiry}~absolute}) \lor (pt_{\mathrm{htlc}, n+1} \land ph_{\mathrm{htlc}, n+1}, \mathrm{on~preimage~of~} h)$ (see Fig. 40, line 8) where $pt$ are public keys for which Alice has the private key, $ph$ are public keys for which Bob has the private key, and $\mathtt{CltvExpiry}$ is the HTLC's absolute timelock.
Now, Alice could spend the output of the commitment transaction corresponding to the HTLC by creating a transaction with an input that uses the disjunct $(pt_{\mathrm{htlc}, n+1}, \mathrm{\mathtt{CltvExpiry}~absolute})$ because the absolute timelock has passed.
Bob holds the HTLC success transaction that was signed by Alice with the private key corresponding to $pt_{\mathrm{htlc}, n+1}$ (Fig. 43, line 13).
If Bob has the preimage for the HTLC, Bob can add the preimage to the HTLC success transaction and can spend the HTLC's output in the commitment transaction using the disjunct $(pt_{\mathrm{htlc}, n+1} \land ph_{\mathrm{htlc}, n+1}, \mathrm{on~preimage~of~} h)$ of the spending method.
However, the HTLC success transaction is also valid without the preimage as it fulfills the conditions of the disjunct $(pt_{\mathrm{htlc}, n+1}, \mathrm{\mathtt{CltvExpiry}~absolute})$ because the HTLC's absolute timelock has passed.
Because Bob published his latest commitment transaction, Alice cannot revoke the transaction and this would result in Bob receiving the amount of the HTLC without releasing (or even without having) the preimage.
One way to correct this problem is to use the possibility that the transaction model of the paper \cite[Section 12]{kiayias_composable_2019-1} allows an output to specify a list of spending conditions and an input spending this output to reference a specific spending condition.
The correction would be to transform the disjunction in Fig. 40, line 8 into a list of spending methods and add the corresponding indices to the inputs in Fig. 43, line 13.
Another way is taken by the Lightning Network's specification which uses in the output's spending method for a timeout the operator \texttt{CHECKLOCKTIMEVERIFY} that verifies that a spending transaction has a certain timelock set (\texttt{locktime}). As Bob's HTLC success transaction has the \texttt{locktime} set to 0, the success transaction cannot fulfill this spending method.
We found this flaw by model checking when we had a similar flaw in a draft of our formalization. We fixed the flaw in our formalization by modeling the \texttt{locktime} field for transactions and adding a validity condition modeling the operator \texttt{CHECKLOCKTIMEVERIFY}.

\end{document}